\documentclass[%
 reprint,
 amsmath,amssymb,
 aps,
]{revtex4-2}

\usepackage{graphicx}% Include figure files
\usepackage{dcolumn}% Align table columns on decimal point
\usepackage{bm}% bold math
\usepackage{hyperref}% add hypertext capabilities
\usepackage{xcolor}
\usepackage[T1]{fontenc} 
\usepackage[utf8]{inputenc}
\begin{document}

\preprint{APS/123-QED}

\title{Learning difficulties among students when applying Amp\`ere-Maxwell’s law and its implications for teaching}

\author{Álvaro Suárez}
%\email{alsua@outlook.com} 
\affiliation{Departamento de Física, Consejo de Formación en Educación, Montevideo, Uruguay}

\author{Arturo C. Marti}
%\email{marti@fisica.edu.uy}
\affiliation{Instituto de F\'{i}sica, Facultad de Ciencias, Universidad de la Rep\'{u}blica, Igu\'{a} 4225, Montevideo, 11200, Uruguay}

\author{Kristina Zuza}
%\email{kristina.zuza@ehu.eus} 
\affiliation{Department of Applied Physics and Donostia Physics Education Research Group, University of the Basque Country (UPV/EHU), San Sebastian 20018, Spain}

\author{Jenaro Guisasola}
%\email{jenaro.guisasola@ehu.eus} 
\affiliation{Donostia Physics Education Research Group, University of the Basque Country (UPV/EHU), San Sebastian 20018, Spain \\ School of Dual Engineering, Institute of Machine Tools (IMH), Elgoibar, Spain}

\date{\today}

\begin{abstract}
We investigate learning difficulties among second-year students on electromagnetism courses when they apply Amp\`ere-Maxwell’s law. Using phenomenography, we analysed written answers from 65 undergraduate physics students to four questions on Amp\`ere’s and Amp\`ere-Maxwell’s laws. We complemented our research by interviewing twelve students. To design the questionnaire, we ran an epistemological analysis of classical electromagnetism  which helped us to identify a set of key essential concepts to understand this theory and guided the definition of learning objectives and to draw up the questions. The results revealed that the students found it hard to recognise the validity framework from Amp\`ere's law and to apply Amp\`ere-Maxwell’s law. They face particular difficulties to recognise the appearance of the displacement current and the relationship between the circulation of the magnetic field and an electric field that is variable over time.
\end{abstract}

\maketitle % title page is now complete

\section{Introduction}
Incorporating J.C. Maxwell’s displacement current into Ampère’s law was a milestone in developing his electromagnetic theory of light. This generalization of Ampere’s law, also known as Ampère-Maxwell’s law, alongside Gauss’s laws for electric and magnetic fields and Faraday’s law, provides the fundamental equations for classic electromagnetic theory known as ‘Maxwell’s laws or equations’. Many studies have been run to address students’ difficulties when applying these laws. This research has mainly focussed on Gauss’s law for electric fields, Ampère’s law and Faraday’s law. 

The difficulties associated with teaching Gauss's law have received considerable attention. In particular, it has been observed that a high percentage of students apply it by rote which generates confusion between concepts included in the law such as the concepts of flux and field, not understanding the meaning of the surface integral operator  \cite{pepper2010,pepper2012,guisasola2003analisis,guisasola2008gauss,Campos2023}, confusion between the charge distribution symmetry and the geometry, and between open and closed surfaces \cite{singh2006,Li_2018}.
In relation to the magnetic fields generated by constant currents, 
{students often show a lack of understanding about the application of  Ampère’s law.
This difficulty leads them to incorrect reasoning, such as considering that circulation can always be expressed as the product of the magnetic field times the length of the curve, supposing that when the net current is null, the magnetic field is also null, or confusing the concepts of circulation, flux and magnetic field \cite{guisasola2008gauss,Campos2023,atasoy2013effect,wallace2010,Campos2023,Hernandez2023}.
Regarding electromagnetic induction, it has been found that students believe that the  \textit{electromotive force} (emf) or the induced current can be generated by a static magnetic field  \cite{Saarelainen_2007,thong2008some,guisasola2013university} and they do not know the difference between a Coulomb electric field and an induced electric field \cite{thong2008some}. When they apply Faraday’s law, they confuse the circuit area with the integration area \cite{zuza2012} and they find it hard to apply Lenz’s law  \cite{Bagno1997}. 

In contrast to these advances, studies dedicated to the Ampère-Maxwell’s law and the displacement current have been more limited, mainly focusing on  analyses about how this topic is presented in university textbooks \cite{pocovi2011corriente,suarez2023} plus teaching approaches and applications \cite{Gauthier1983,Reich,karam2014,moreno2020ensenanza,milsom2020untold,Mungan}.
Interestingly, Ampère-Maxwell’s law, and how the displacement current is introduced, are the subject of frequent discussion and controversy within the scientific community. Questions have been raised regarding the role of charge conservation in Ampère-Maxwell’s law 
\cite{Zapolsky,Wolsky_2015}, the causal nature of this law 
\cite{jefimenko2004presenting,hill2011reanalyzing}, its equivalence to Biot-Savart’s law \cite{Weber1989,buschauer2013derivation} if the displacement current represents a true current and if it generates a magnetic field \cite{Rosser1976,Roche_1998,Jackson1999,french2000maxwell,Heras2011,landini2014,Hyodo_2022}. 

Despite progress in PER relating to students’ learning difficulties when applying Maxwell’s equations, a lack of attention to students’ comprehension of Ampère-Maxwell’s law can be identified. Motivated by this situation, our main research question is: “What learning difficulties do undergraduate students face when applying Ampère-Maxwell’s law?” and, more specifically,
\begin{itemize}
    \item RQ1. Do undergraduate students understand the limitations of Ampère’s law?
    \item RQ2. How do undergraduate students understand and apply Ampère-Maxwell’s law?  
\end{itemize}
To address these questions, our research began with an epistemological analysis of how classic electromagnetism developed and later developed into Ampère-Maxwell’s law, which helped us to identify a set of key concepts to understand it.  These key concepts guided us as we defined learning goals for a curriculum on Ampère-Maxwell’s law in electromagnetism courses for Physics and Engineering undergraduates. Working from the learning objectives, we designed a questionnaire encompassing four situations related to Ampère-Maxwell’s law and displacement current. We gave this questionnaire to second-year university students taking electromagnetism courses. We assessed their answers using a phenomenographic approach and ran interviews to delve deeper into their learning. Analysing the results helped us to identify the learning difficulties students face when applying Ampere-Maxwell’s law.

Below, in section \ref{sectionII}, we present the epistemological development of classical electromagnetism and its development into Ampère-Maxwell’s law. In section \ref{sectionIII}, we describe the study context and explain the methodology. Section \ref{SectionIV} focuses on presenting the results, and we discuss these results in section \ref{SectionV}. Finally, we present the conclusions from our work and its implications for teaching in section \ref{SectionVI}. 

\section{\label{sectionII}Epistemological development of classical electromagnetism and its development into Ampere-Maxwell’s law} 

The development of classical electromagnetic theory, as we know it today, plus the formulation of its fundamental laws, Maxwell’s equations, was the result of a laborious process during the 19th century with several epistemological and ontological obstacles. The figure of James Clerk Maxwell came to the forefront halfway through that century. Influenced by Michael Faraday’s ideas on fields, he devoted two decades to building the foundations of a theory of electromagnetic fields. While maturing his concept, Maxwell incorporated an element which proved to be essential when unifying light and electromagnetism: the displacement current. Exploring how his ideas developed helps us to understand the key concepts associated with Ampère-Maxwell’s law and the displacement current.

In 1855, Maxwell published \textit{“On Faraday's Lines of Force”}. In this work, he uses the idea of continual transmission of electric and magnetic force, as Faraday imagined, and he considers lines of force as states of a mechanical ether, as conceived by William Thomson. Working from these ideas and using analogies for the incompressible movement of a fluid, he developed a mathematical formulation for Faraday’s lines of force  \cite{berkson2014fields,darrigol2003electrodynamics}. This work presents an initial contribution to developing Ampere-Maxwell’s law. When analysing the continuity equation for the conduction current, he realised that his first drafts of an electromagnetic theory would not allow him to tackle problems related to open circuits \cite{Bork1963}. In this context, Maxwell stated: 
\begin{quote}
    \textit{“Our investigations are therefore for the present limited to closed currents; and we know little of the magnetic effects of any currents which are not closed”} \cite[p~195]{maxwell1890scientific}
\end{quote}

A second important factor is the introduction of the concept of displacement current in \textit{“On Physical Lines of Force”} (1861), by establishing a connection between electrical conduction in conductors and electric displacement in insulators \cite{karam2014}. Maxwell explained this concept in the following way: 
\begin{quote}
    \textit{“Bodies which do not permit a current of electricity to flow through them are called insulators. But though electricity does not flow through them, the electrical effects are propagated through them […] Here then we have two independent qualities of bodies, one by which they allow of the passage of electricity through them, and the other by which they allow of electrical action being transmitted through them without any electricity being allowed to pass. […] The effect of this action on the whole dielectric mass is to produce a general displacement of the electricity in a certain direction. This displacement does not amount to a current, because when it has attained a certain value it remains constant, but it is the commencement of a current, and its variations constitute currents in the positive or negative direction, according as the displacement is increasing or diminishing”} \cite[pp~144-145]{maxwell1890scientific}
\end{quote}
By introducing the displacement current, Maxwell developed a version of the continuity equation, similar to what we use nowadays  \cite[p~496]{maxwell1890scientific}. At this historical point of time, Maxwell considered the displacement current to be a component of the conduction current \cite{darrigol2003electrodynamics}.

One of the most outstanding achievements of \textit{“On Physical Lines of Force”} was identifying light as an electromagnetic radiation phenomenon, where Maxwell managed to deduce the electromagnetic waves velocity \cite{nersessian2012} and proposed that \textit{“light consists in the transverse undulations of the same medium which is the cause of electric and magnetic phenomena”} \cite[p~500]{maxwell1890scientific}. In his deduction process, Maxwell was obliged to use auxiliary hypotheses related to the mechanism that he conceived for the electromagnetic field to determine the wave velocity, as he could not find a wave equation for the electromagnetic field. Having reached this point and aware of the difficulties associated with his mechanical model of the ether, he decided to separate it from his electromagnetic theory. He intended to obtain field equations disassociated from any specific model of the ether and deduce a wave equation that would make it possible to predict the speed of light. Consequently, in 1864, he published \textit{“A Dynamical Theory of the Electromagnetic Field”}, presenting eight equations to describe the electromagnetic field and an electromagnetic theory of light that might be subject to experimental tests \cite{harman1982energy,berkson2014fields}. In this new work, Maxwell addressed the challenge of how to interpret his set of equations. This saw the emergence of the analytical interpretation, which states that it is the electromagnetic field which acts directly on the matter at any point in space. The matter and the field thereby became separate entities and thereby lost the meaning of the adjacent force mechanisms imagined by Faraday \cite{berkson2014fields}.

Another important contribution in clarifying the concept of displacement current is presented by Maxwell stating that  
\begin{quote}
    \textit{“Electrical displacement consists in the opposite electrification of the sides of a molecule or particle of a body which may or may not be accompanied with transmission through the body… The variations of the electrical displacement must be added to the currents p, q, r to get the total motion of electricity…”} \cite[p~554]{maxwell1890scientific}
\end{quote}
This leads to two fundamental conclusions. Firstly, as opposed to what is laid out in \textit{“On Physical Lines of Force”}, Maxwell now considers that the displacement current is a different type of current which contributes to the total current \cite{chalmers1975maxwell,darrigol2003electrodynamics}. Secondly, the concept of electrical displacement, as conceived by Maxwell, is what we know today as the polarisation vector \cite{arthur2009elementary}.

A third epistemological leap in developing the theory happened when Maxwell argued the need to include the displacement current so that Ampere’s law remained general. In 1873, he published his definitive work, \textit{“A Treatise on Electricity and Magnetism”}, which presented his entire electromagnetic theory in detail. In this work, he clearly expresses his position on the nature of electric current from the temporary variation of the electric displacement, by stating that: 
\begin{quote}
\textit{“The current produces magnetic phenomena in its neighbourhood
  […] We have reason for believing that even when there is no proper
  conduction, but merely a variation of electric displacement, as in
  the glass of a Leyden jar during charge or discharge, the magnetic
  effect of the electric movement is precisely the same.”}
\cite[pp~144-145]{maxwell1873treatise}
\end{quote}
He then refers to the continuity equation and demonstrates how including the displacement current leads him to consider that all the circuits are closed, as long as both types of current are considered. He states: 
\begin{quote}
\textit{“By differentiating the equations E (what we call Ampère-Maxwell’s law nowadays) with respect to x, y and z respectively, and adding the results, we obtain the equation du/dx+dv/dy+dw/dz=0, which indicates that the current whose components are u, v, w is subject to the condition of motion of an incompressible fluid, and that it must necessarily flow in closed circuits. This equation is true only if we take u, v, and was the components of that electric flow which is due to the variation of electric displacement as well as to true conduction. We have very little experimental evidence relating to the direct electromagnetic action of currents due to the variation of electric displacement in dielectrics, but the extreme difficulty of reconciling the laws of electromagnetism with the existence of electric currents which are not closed is one reason among many why we must admit the existence of transient currents due to the variation of displacement. Their importance will be seen when we come to the electromagnetic theory of light.”}
\cite[p~233]{maxwell1873treatise}
\end{quote}
This last sentence clearly shows how important the displacement current was to Maxwell and how it became a fundamental part of his electromagnetic theory of light. Maxwell recognised that it was essential to include the displacement current to reconcile the laws of electromagnetism with the existence of electric currents that are not closed, and this played a crucial role in his formulation of the electromagnetic theory of light.

Years after Maxwell’s death, Heinrich Hertz performed a series of experiments in 1887 which demonstrated the existence of electromagnetic waves. These experiments heralded important progress in confirming the electromagnetic field theories and helped to raise the status of Maxwell’s field theory. In 1892, Hendrik Lorentz published \textit{“La théorie électromagnétique de Maxwell et son application aux corps mouvants”}, where he proposed the existence of the electron and introduced the idea of an ether completely at rest, making an analytical interpretation of the field equations the only possible approach \cite{berkson2014fields}. Years later, when Einstein’s special theory of relativity was published, which brought with it the fall into disuse of the concept of ether, the \textit{“analytical interpretation”} for Maxwell’s equations prevailed.

Working from the description of the main obstacles and challenges involved in consolidating the classical electromagnetic theory, particularly including the displacement current as a fundamental concept to develop the electromagnetic theory of light, we can identify a set of key concepts to be considered when addressing the displacement current and Ampere-Maxwell’s law at an undergraduate physics introductory level.
\begin{itemize}
    \item KC1. Ampère’s law is only valid in stationary circuits where the current is closed (Maxwell 1855, On Faraday's Lines of Force).
    \item KC2. The validity of charge conservation implies the need to include the displacement current (Maxwell, 1861, On Physical Lines of Force).
    \item KC3. Ampère-Maxwell’s law is general and must include conduction and displacement currents to calculate the magnetic field circulation (Maxwell, 1873, A Treatise on Electricity and Magnetism).
\end{itemize}

The conceptual and epistemological clarification provided by the key concepts helps us to explicitly define learning goals regarding Ampère-Maxwell’s law for undergraduate Physics and Engineering electromagnetism courses, as shown in table \ref{claves}.

\begin{table*}
 \caption{\label{claves} Conceptual keys and learning goals.}
\begin{tabular}{|c|p{14 cm}|}  
\hline
\textit{Conceptual keys}
 & \textit{Learning objectives for Ampère-Maxwell’s law.}\\ \hline
KC1 
 &\begin{itemize}
    \item LO1. Identify the current that crosses a surface bound by a closed curve.
    \item LO2. Recognise situations where, by using different surfaces bound by the same curve, contradictory results are obtained for magnetic field circulation.
\end{itemize}
\\ \hline
KC2 and KC3
 &\begin{itemize}
    \item LO3. Recognise that displacement currents exist and apply the concept correctly in Ampère-Maxwell’s law.
    \item LO4. Understand that a variable electric field is associated with a magnetic field.
    \item LO5. Appropriately connect the rate of change of the electric field to magnetic field circulation.
\end{itemize}
\\ \hline
\end{tabular}
\end{table*}

\section{\label{sectionIII}METHODOLOGY} 

In this study, we decided to research students’ learning difficulties when applying Ampère-Maxwell’s law. We specifically focused on exploring whether students properly understand the limitations of Ampère’s law and how to interpret and apply Ampere-Maxwell’s law in different contexts. We thereby designed a questionnaire,
transcribed in the appendix \ref{apppendix}, that was given to undergraduate students on electromagnetism courses and used it as a written test, using phenomenographic methodology to analyse the answers\cite{marton1981phenomenography}.
  Additionally, and to delve deeper into student reasoning, we asked these questions in interviews. 
A description of the study participants, the assessment instrument and the procedures used to analyse the compiled data are presented below.

\subsection{\label{IIIA}Participants}
The sample selected second-year Physics undergraduates, distributed as follows: 58 came from the University of the Basque Country (UPV/EHU) and 19 from the University of the Republic of Uruguay (UDELAR). All participants in the research had previously passed introductory physics courses with similar curricula and reference bibliographies \cite{tipler2007physics,young2019university}.

The data were compiled using questionnaires given as written tests. Furthermore, 12 students were interviewed on the questions set in the study. Before taking part in the research, all students received classes on Ampère’s and Ampère-Maxwell’s law. The electromagnetism courses at both universities lasted 15 weeks, with 4 hours of theory classes and 2 hours of problem-solving per week. Both courses use a standard teaching approach (lectures and classes where the teacher demonstrates how to solve problems) and the level was based on the recommended bibliography, derived from the principal textbooks used in introductory physics courses in the United States, Latin America and Spain \cite{suarez2023}. All the courses were taught by experienced university professors. 

\subsection{\label{IIIB}Questionnaire}

To devise the research instrument, we used key concepts derived from the epistemological development of classic electromagnetism and its development into Ampère-Maxwell’s law  (section \ref{sectionII}). Based on these key concepts and considering the educational context, which addresses the students’ academic level, their prior knowledge and the reference bibliography, we defined the learning objectives associated with appropriate comprehension of Ampère-Maxwell’s law \cite{Guisasola2017,zuza2020towards,guisasola2021,Guisasola2023AIP}.These goals are closely bound to the research questions (RQ1 and RQ2) and they played a fundamental role in creating the research instrument, which consists of a questionnaire comprising four situations. 

Question 1 was designed to evaluate the students’ comprehension regarding the limitations of Ampère’s law. It introduces a scenario which shows two charged conducting spheres with opposite charges, connected together using a conducting wire as shown in Fig. \ref{cuestion1}. Two surfaces, $S_{1}$ and $S_{2}$, are presented, bound by a curve \textit{C} and the students were asked to indicate how they would apply Ampère’s law to each surface. This situation is familiar to the students in an academic context. The students are expected to be able to recognise that when applying Ampère’s law to the surfaces $S_{1}$ and $S_{2}$, contradictory results are obtained and that this is directly related to its validity framework (LO1 and LO2).

\begin{figure}[htbp!]
\centerline{\includegraphics[width = 0.7\columnwidth]{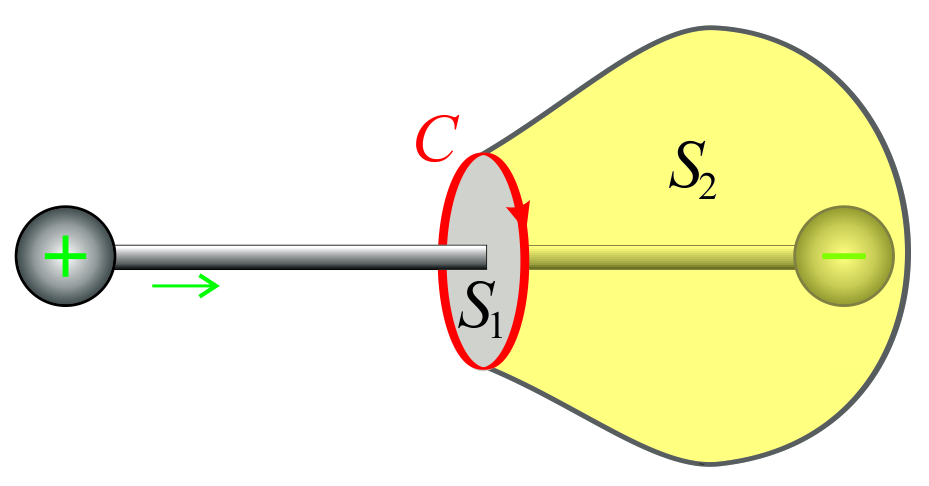}}
\caption{Scheme included in question 1.}
\label{cuestion1}
\end{figure}

Question 2 presents a scenario with two charged conducting spheres with opposite charges, connected using a conducting wire. It shows a curve \textit{C} and a surface \textit{S} (see Fig. \ref{cuestion2}), and asks whether, when applying Ampere-Maxwell’s law to calculate the magnetic field circulation along curve \textit{C}, this would obtain a result which is greater than, less than or equal to what was predicted by Ampère’s law. This question aims to investigate whether students understand the conditions in which a displacement current appears and how they consider it when applying Ampère-Maxwell’s law (LO3). They are expected to recognise that when connecting the spheres using a wire, a conduction current is established that decreases as the spheres discharge, resulting in a decrease in the electric field at each point in space and consequently, in the electric flux through surface \textit{S}. Consequently, a displacement current appears in the opposite direction to conduction, which leads to lower magnetic field circulation than predicted by Ampere’s law.

\begin{figure}[htbp!]
\centerline{\includegraphics[width = 0.6\columnwidth]{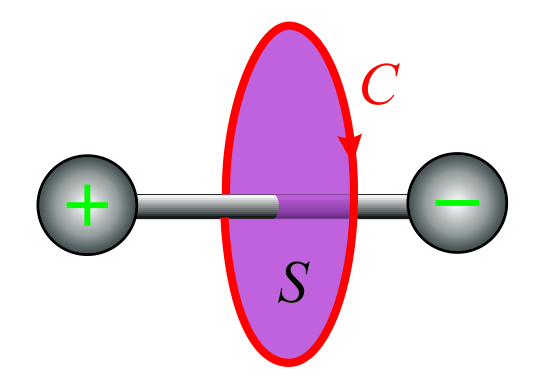}}
\caption{Scheme included in question 2.}
\label{cuestion2}
\end{figure}

In question 3  students must explain the presence of a magnetic field where the field lines have cylindrical symmetry, as shown in Fig. \ref{cuestion3}, indicating that it was detected in a empty region. The aim of this question is to investigate whether the students can recognise that the existence of a magnetic field in a particular region of empty space implies the simultaneous presence of a variable electric field over time (LO4).

\begin{figure}[htbp!]
\centerline{\includegraphics[width = 0.8\columnwidth]{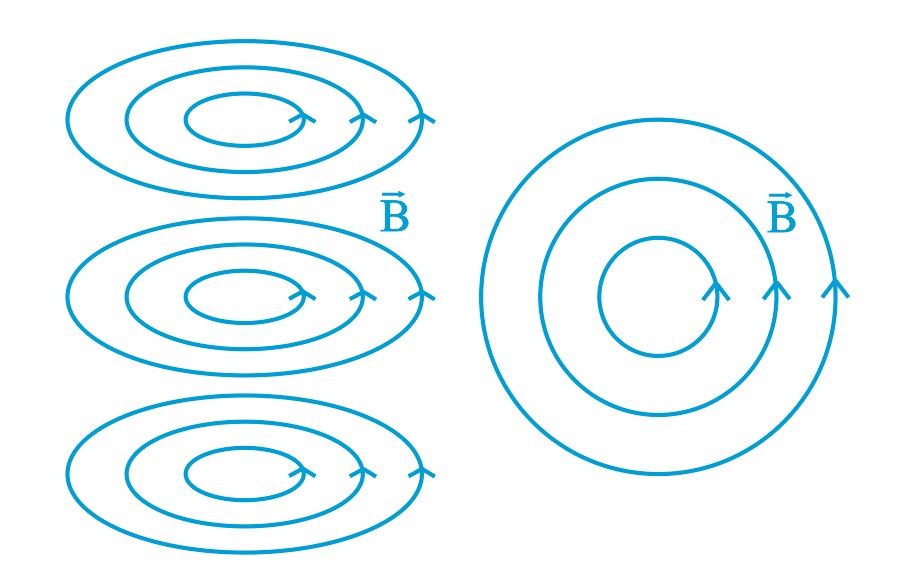}}
\caption{Scheme included in question  3.}
\label{cuestion3}
\end{figure}

Question 4, shown in Fig. \ref{cuestion4},  present an  electric field pointing into the page, a closed curve \textit{C} and a graph of the electric field over time. The students must sketch the graph of the magnetic field circulation ($\oint_{C} \vec{B} \cdot \overrightarrow{d l}$) along curve \textit{C}  over time. 
The aim of this question is to assess whether they understand the relationship between the rate of change of the electric and the magnetic field circulation. It is expected that they will be capable of using the information provided in the graph to deduce the shape of the magnetic field circulation graph (LO5).

\begin{figure}[htbp!]
\centerline{\includegraphics[width = 1.0\columnwidth]{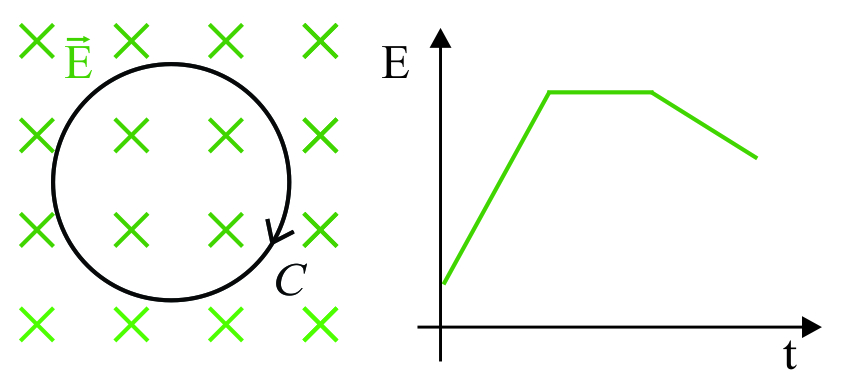}}
\caption{Scheme included in question 4.}
\label{cuestion4}
\end{figure}

The questionnaire content was subject to a validation process requiring physics teachers and PER experts to work together  \cite{cohen2007research}, followed by a pilot study with undergraduate Physics students from an electromagnetism course at the UDELAR. The pilot study aimed to ensure that the students understood the meaning of the various questions and that the answers matched the learning objectives for which they had been designed.

Table \ref{objetivos} shows the relationship between the research questions, learning objectives and the different questions.

\begin{table*}
 \caption{\label{objetivos} Learning objectives and their relationship with the research objectives and the questionnaire.}
\begin{tabular}{|p{5 cm}|p{9 cm}|c|}  
\hline
\textit{Research questions}
 & \textit{Learning objectives }& \textit{Questions}\\ 
 \hline
RQ1. Do they understand the limitations of Ampère’s law?
 & LO1. Identify the current that crosses a surface bound by a closed curve. & Q1 
\\ 
\hline
 & LO2. Recognise situations where, by using different surfaces bound by the same curve, contradictory results are obtained for magnetic field circulation. & Q1 
\\ 
\hline
RQ2. How do they understand and apply Ampère-Maxwell’s law?
 & LO3. Recognise that displacement currents exist and apply the concept correctly in Ampère-Maxwell’s law.
 & Q2 
\\ 
\hline
 & LO4. Understand that a variable electric field is associated with a magnetic field.
 & Q3 
\\ 
\hline
 & LO5. Appropriately connect the rate of change of the electric field to magnetic field circulation.
 & Q4 
\\ 
\hline
\end{tabular}
\end{table*}

The questions were designed to investigate how the students understand and apply Ampère-Maxwell’s law in different contexts. By looking at different situations, we can assess whether the students have a deeper understanding of Ampère-Maxwell’s law. According to Howard Gardner \cite[pp~13-14]{gardner2011}, \textit{“Genuine understanding is most likely to emerge, and be apparent to others, if people possess a number of ways of representing knowledge of a concept or skill and can move readily back and forth among these forms of knowing.”} Analysing the students’ answers to the various questions provides valuable information on how they interpret Ampère-Maxwell’s law and how they apply it in different situations, thereby identifying possible learning difficulties.

\subsection{\label{IIIC}Data collection and analysis}

We used phenomenographic analysis to analyse the written tests  \cite{Guisasola2023PR}. This method is specifically designed to investigate the different ways that people experience and understand a particular phenomenon. Phenomenography is particularly useful to recognise and appreciate the singularity of individual experiences and understanding, identifying common experiences and conceptions that can be grouped together to describe a collective understanding \cite{marton1981phenomenography,Guisasola2023PR}. This method has proven to be valuable in physics education research, particularly when analysing written open questions, to identify how each student reasons and their learning difficulties \cite{Guisasola2023PR}. 

The analysis process was carried out by two of the researchers and began by studying the answers from an initial group of 20 students. Thanks to consensus among the researchers, we identified emerging categories that met certain essential characteristics in the context of phenomenographic analysis. These categories had to reflect variations in the students’ understanding, revealing different aspects of how they understood the topic, organised logically by establishing hierarchical, structurally-inclusive relationships. In addition, the set of categories had to be as small as possible, representing the critical variation of the experiences observed in the data  \cite{marton1997booth,Guisasola2023PR}. Each researcher then analysed the remaining answers independently and finally, we compared the results. To assess the reliability of our analysis, we used the Cohen kappa coefficient, a measurement that quantifies the degree of agreement between assessors, considering the possibility of random coincidences. We obtained an average value of 0.95. A kappa value over 0.80 is considered to indicate significant agreement \cite{banerjee1999beyond}.

After analyzing the written tests, we proceeded to conduct semi-structured group interviews. In total, five interviews were conducted, three with groups of two students and another two with groups of three participants. The purpose of these group interviews was to encourage discussions and obtain enriching perspectives that might not emerge in individual interviews \cite{cohen2007research}. The interviews includes the same questions as the questionnaire and the participants were selected so that they represent a wide variety of academic levels, with the purpose of guaranteeing the external validity of the results  \cite{otero2009}. During the interviews, we asked the students to solve the questions out loud and explain their reasoning and the procedures they used. The interviews were run by one of the researchers who, when they thought it necessary, asked the students to clarify or expand on their explanations, to ensure appropriate comprehension of what the interviewees were expressing. All the interviews were recorded in audio format and transcribed for analysis. When presenting the results, fictitious names were used to ensure anonymity.

\section{\label{SectionIV}RESULTS} 

Below, we present a detailed description of the most relevant results derived from the analysis of the written tests and interviews in each of the questions.

\subsection{\label{SectionIVA}Question 1}

When analysing the students’ answers, we identified a set of categories that reflect different types of explanations (see table \ref{Tabla cuestión 1}).  

\begin{table*}
 \caption{\label{Tabla cuestión 1} Categories of answers and their percentages for question 1.}
\begin{tabular}{|c|p{12,9 cm}|c|}  
\hline
\textit{Category}
 & \textit{Description of the category} & \textit{\%}\\ \hline
A & The students understand that the two surfaces are bound by the same curve and that the same result should be obtained. They recognise the limitation of Ampere’s law. & 21,6 \\ \hline
B & The students demonstrate incomplete comprehension of Ampere’s law. They apply Ampere’s law to $S_{1}$ but they do not understand the law’s limitation for surface $S_{2}$. & 24,6 \\ \hline
B1 & They state that there is a problem with $S_{2}$. & 15,4 \\ \hline
B2 & They confuse flux with circulation in $S_{2}$. & 9,2 \\ \hline
C & The students do not understand the meaning of Ampere’s law. & 35,4 \\ \hline
C1 & Wrong or incoherent descriptions when solving the magnetic field integral in $S_{1}$. & 21,6 \\ \hline
C2 & They confuse flux with circulation when solving the magnetic field integral equation in $S_{1}$. & 13,8 \\ \hline
D & Incoherent. & 9,2 \\ \hline
E & They do not answer. & 9,2 \\ \hline
\end{tabular}
\end{table*}

Category A (21,6\%) includes answers that correctly understand Ampère’s law and its limitations. The students in this category recognise that the two surfaces are bound by the same curve and consequently that the same result should be obtained when applying Ampère’s law. They then spot a contradiction and, as a result, they suggest using the full form of the law. Let’s look at a few examples. 

\begin{quote}
\textit{A - “In this problem, when applying Ampère’s law, we see the error and the need for Maxwell’s laws. By applying Ampère’s law on $S_{1}$ and $S_{2}$, we supposedly should obtain the same result because both surfaces have the same shape. However, this is not true, $\oint \vec{B} \cdot \overrightarrow{d l}=\mu_{0} I_{c}$ is Ampère’s law. The current that is generated when connecting the two charges crosses $S_{1}$ but does not cross $S_{2}$.”} (UPV/EHU-29)
\end{quote}

\begin{quote}
\textit{A-  “For surface $S_{1}$ there is a conduction current but not for surface $S_{2}$. Consequently, a displacement current should be considered through the latter, given that the enclosed current is the same in both. Ampère’s law must be applied considering a displacement current and a conduction current to complete it.”} (UDELAR - 1)
\end{quote}

In category B (24,6\%), there were students who demonstrate an incomplete degree of comprehension of the physics principles involved. Within this category, we identified two sub-categories: B1, where the students apply Ampère’s law to surface $S_{1}$ and state that there is a problem with $S_{2}$ but they do not solve it, and B2, where the students apply Ampère’s law on $S_{1}$, but confuse the concepts of flux and circulation when analysing what happens on $S_{2}$. Let’s look at a few examples from each sub-category. 

\begin{quote}
\textit{B1 - “I think that we can use the equation $\oint \vec{B} \cdot \overrightarrow{d l}=\mu_{0} I_{c}$. For the case of $S_{1}$, $B \int 2 \pi r d r=\mu_{0} I \rightarrow B=$ $\frac{\mu_{0} I}{2 \pi R}$. In the case of $S_{2}$, I don’t know how to solve it.”} (UPV/EHU-15)
\end{quote}

\begin{quote}
\textit{B1 - “I would apply Ampère’s law for $S_{2}$ as specified below  $\oint \vec{B} \cdot \overrightarrow{d l}=\mu_{0} I$. No current passes through $S_{2}$, so there will be no B.”} (UPV/EHU-27)
\end{quote}

\begin{quote}
\textit{B2 - “For surface $S_{1}$, we apply $\oint \vec{B} \cdot \overrightarrow{d l}=B 2 \pi r$ to calculate the magnetic field from here. For $S_{2}$, as the vector $\hat{n}$ is perpendicular to B, it will be cancelled out.”} (UPV/EHU-26)
\end{quote}

\begin{quote}
\textit{B2 - “$S_{1} \rightarrow B 2 \pi r=\mu_{0} I \rightarrow B=\frac{\mu_{0} I}{2 \pi r}$. The $S_{2}$ surface is not closed so Ampère’s law cannot be applied.”} (UPV/EHU-32)
\end{quote}

In category B1, the students used Ampère’s law appropriately for surface $S_{1}$, but they do not know how to apply the law for the surface $S_{2}$ (UPV/EHU-15) or they use arguments based on the current, supposing that if the current is null, the magnetic field is also null (UPV/EHU-27). Despite managing to apply Ampère’s law correctly on surface $S_{1}$, students UPV/EHU-26 and UPV/EHU-32 in category B2, confuse the concepts of flux and circulation when trying to apply it to surface $S_{2}$.  

In category C, we included answers from students who state that they know about magnetic field circulation in relation to Ampère’s law, but they are not capable of recognising any contradiction between the surfaces. They represent around 35\% of the total. We divided this category into two sub-categories: C1, where the students solve or try to solve the magnetic field integral equation on $S_{1}$, and C2, where the students confuse the concepts of flux and circulation when trying to apply Ampère’s law. Let’s look at a few examples from each sub-category. 

\begin{quote}
\textit{C1 - “Ampère’s law $\oint \vec{B} \cdot \overrightarrow{d l}=\mu_{0} I_{\text {encerrada. }}$. I think that Ampère’s law is independent from the surface. Therefore, you would deal with it in the same way, as the enclosed current is the same, the result would be the same.”} (UPV/EHU-34)
\end{quote}

\begin{quote}
\textit{C1 - “Firstly, we would apply the following formula $\oint \vec{B} \cdot \overrightarrow{d l}$ and to be able to use the data I’ve been given, I would take the necessary vectors and the radii from the two surfaces where we would obtain values for $R$ and $\alpha$.”} (UPV/EHU-8)
\end{quote}

\begin{quote}
\textit{C2 - “In $S_{1}$, the angle with the wire is always the same, but in the case of $S_{2}$, the angle changes and consequently cannot be calculated in the same way for both surfaces.”} (UPV/EHU -16)
\end{quote}

\begin{quote}
\textit{C2 - “The magnetic flux is zero in $S_{1}$ because I is perpendicular to the surface. I don’t know what to do with $S_{2}$”} (UDELAR-5)
\end{quote}

In the C1 category, the UPV/EHU-34 student’s reasoning is based on the current which passes through the curve and concludes that the magnetic field circulation does not change, because the current bound by the curve is the same. They do not argue according to the physical meaning of the concepts implicated in Ampère’s law. This type of response demonstrates conceptual difficulties to identify the current intensity which crosses different surfaces bound by the same curve. In category C2, the UDELAR-5 student finds it hard to understand the concept of circulation, confusing it with field flux.

In the reasoning used in both categories B and C, we can see how introducing the curved surface elicits different answers regarding magnetic field circulation. This suggests inadequate comprehension of the concept of circulation and the different terms in Ampère’s law. To deepen the analysis of this lack of comprehension, we examined extracts from two interviews which show how the context highlights the difficulties regarding the concept of magnetic field circulation and Ampère’s law. 

\begin{quote}
    \textit{Interviewer}: Could you apply Ampère’s law to curve \textit{C} and surface $S_{2}$? \\\textit{Fabiana}: I’m not sure that I can apply Ampère’s law in $S_{2}$. \\\textit{Interviewer}: Why?\\\textit{Fabiana}: Firstly, it is not completely closed, it is open, it is like a hollow and it is surrounding the charge. If I had to say, I’d say no.\\\textit{Interviewer}: And why would you say no? \\\textit{Fabiana}: You can’t apply Ampère’s law in this part (referring to surface $S_{2}$) because the field lies around the conductor this way. The field comes as surface $S_{1}$ (she draws the magnetic field lines concentric to the wire on curve \textit{C}). So, there is no magnetic field in $S_{2}$ and Ampère’s law cannot be applied.
\end{quote}

Fabiana’s arguments demonstrate her lack of comprehension regarding the concept of circulation, due to the shape of the surface $S_{2}$. Fabiana states that there is no magnetic field on surface $S_{2}$, which stops her applying Ampère’s law on this surface. She does not recognise that the magnetic field circulation is calculated over closed curve \textit{C} and that this curve delimits both surfaces $S_{1}$ and  $S_{2}$. 

When we asked another student how he would apply Ampère’s law on surfaces $S_{1}$ and $S_{2}$, he answered as follows:
\begin{quote}
  \textit{Federico}: I separate $S_{1}$ in a curve. The only thing that I have to determine for surface 2 is surface S. What you can do in $S_{2}$ is add, for example, you have a curve, another curve, another curve, another curve, another curve, and you add them as you go along. And there you have the sum of curves. And you are building, the only thing you are doing is breaking up that surface into many curves (see Fig. \ref{entrevista federico}).
  Then you separate the surface into curves, and you add it all together, with some mathematical magic, and it should work.\\\textit{Interviewer}: Would it give you the same or a different result as applying Ampère’s law on both surfaces?\\\textit{Federico}: It should give the same result because we are talking about the net current that passes through the centre.
\end{quote}

\begin{figure}[htbp!]
\centerline{\includegraphics[width = 0.4\columnwidth]{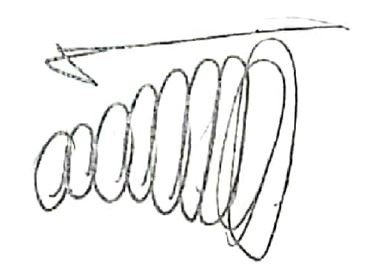}}
\caption{Hand-drawn sketch given by Federico during the interview.}
\label{entrevista federico}
\end{figure}

Federico does not understand the meaning of magnetic field circulation and invents a way of calculating it using a sum of curves. This argument shows a lack of understanding that the current intensity in Ampère’s law is what crosses the surface bound by the curve. In the end, he decides to reason based on the current: if it is crossed by the same current, then circulation must be the same. These two interviews show how the context provided by the curved surface reveals a lack of comprehension regarding Ampère’s law and the concept of magnetic field circulation.

\subsection{\label{SectionIVB}Question 2}

Working from the analysis of the students’ explanation, we identified several categories to classify the different types of reasoning. Table \ref{Tabla cuestión 2} presents the different categories alongside their respective percentages.  

\begin{table*}
 \caption{\label{Tabla cuestión 2} Answer categories and their percentages for question 2.}
\begin{tabular}{|c|p{12,5 cm}|c|}  
\hline
\textit{Category}
 & \textit{Description of the category} & \textit{\%}\\ \hline
A & Explanations based on arguments that recognise the existence of the displacement current and take it into account to determine the magnetic field circulation. & 12,3 \\ \hline
B & Explanations that do not consider the displacement current, but that do recognise that the conduction current changes over time and consequently, they consider that Ampère’s law is valid to solve the question. & 12,3 \\ \hline
C & Explanations that consider the conduction current as constant. & 43,1 \\ \hline
C1 & They do not recognise the variation in electric flow, they apply Ampère’s law or Ampère-Maxwell’s law. & 27,7\\ \hline
C2 & They test ‘ad-hoc’ explanations. & 15,4\\ \hline
D & Incoherent. & 6,2 \\ \hline
E & They do not answer. & 26,1 \\ \hline
\end{tabular}
\end{table*}

Category A (12,3\%) includes the students who understand the phenomenon properly, identifying that the value of the magnetic field circulation differs from what was predicted by Ampère’s law. This category includes correct answers and answers that, when assessing the contribution of the displacement current, consider that it has the same sign as the conduction current. Representative examples of this category are presented below.

\begin{quote}
\textit{“A - “When we connect the two charges, the potential difference drops to null, so $\frac{d \Phi_{E}}{d t}$ is negative. In Ampère-Maxwell’s law, by including this variable the magnetic field circulation will be lower as in this formula, the effects of the flux are subtracted from the effects of the current.”} (UPV/EHU-5)
\end{quote}

\begin{quote}
\textit{A - “A greater result, as with Ampère we had $B=\frac{\mu_{0} I}{2 \pi r}$ and with Maxwell $\oint \vec{B} \cdot \overrightarrow{d l}=$ $\mu_{0}\left(I_{c}+\varepsilon_{0} \frac{d}{d t} \int \vec{E} \cdot \overrightarrow{d A}\right)$, as there is an electric field that crosses the surface which is varying over time.”} (UPV/EHU-30)
\end{quote}

Although the students in the previous examples get opposing results, they recognise that there is an electric field or an electric flux that varies over time and that this implies the presence of a displacement current which contributes to the magnetic field circulation along curve C. Another more qualitative argument used by the students to recognise the existence of the displacement current was to identify its cause in the temporal change in the conduction current. This argument is clearly demonstrated in the students’ discussions during the interviews. Below, we present an extract from one of the interviews where two students exchange ideas on whether the displacement current exists in the circuit and its influence on magnetic field circulation.

\begin{quote}
    \textit{Camila}: The quantity of charge that is going to have crossed is not always going to be the same. It is going to decrease until it balances out. \\\textit{Andrés}: You have two connected charges, there is going to be an electric current and clearly afterwards when they balance out, there is no longer going to be a potential difference.\\\textit{Camila}: However, that current is a variable conduction current.  \\\textit{Andrés}: Yes, that’s right. \\\textit{Camila}: Yes, it is a conduction current and variable. \\\textit{Interviewer}: So, is there a displacement current?\\\textit{Camila}: Yes, because the conduction current varies.\\\textit{Interviewer}: And what happens with the magnetic field circulation?\\\textit{Camila}: If you use Ampère, I mean, I don’t know, I suppose that you will assume a direct current, in other words, like a direct conduction current and as it is actually decreasing... In relation to what was predicted by Ampère’s law, it would be greater (the magnetic field circulation) than what you had before. I mean, because of what I’m saying now.\\\textit{Andrés}: Actually, for me it came out higher and you were missing something that must have been the displacement current, which is what Maxwell contributes.
\end{quote}

The answers reveal that they recognise that the conduction current decreases over time because the spheres are discharging. In this context, Camila concludes that there is a displacement current. However, neither of them manages to identify the contribution of the displacement current properly, possibly because of their difficulty to associate it with the variable electric field.
 
We include any answers that do not recognise the displacement current in category B (12,3\%). They consider that the conduction current varies over time and that Ampère’s law is valid for the phenomenon being analysed. We present examples of this category below.

\begin{quote}
\textit{B - “$\oint_{C} \vec{B} \cdot \overrightarrow{d l}$ should always be equal to $\mu_{0} I_{e n c}$, what happens is that the time  $I_{\text {enc }}$ is constantly decreasing until it reaches 0.”} (UDELAR-6)
\end{quote}

\begin{quote}
\textit{B - “Because the only phenomenon that appears is an electric current generated by a potential difference, Ampère’s Law is met as shown, making the line integral equal to the constant due to the intensity I(t).”} (UDELAR-3)
\end{quote}

In this case, the answers reveal that they recognise that the conduction current is variable. They then apply Ampère’s law and conclude that magnetic field circulation is a function that varies over time in proportion to the conduction current. The interviews provide the opportunity to delve into the possible causes of this reasoning. In the following dialogue, three students exchange ideas on solving the question. 

\begin{quote}
    \textit{Alejandro}: When was there displacement current?\\\textit{Diana}: When you don’t have a wire. One example of that is when you have a capacitor. You had a capacitor and a surface in between and because actually something was happening that has the effect as if it were a current. But really, there was no conductor that joined the capacitor plates.\\\textit{Martín}: Of course, what happened was that if you work it out, Ampère’s law gives you the answer until you come across a curve, a surface that of course you make like a little pouch.\\\textit{Interviewer}: Could you draw the little pouch you’re talking about?\\\textit{Martín}: There you have the wire and the capacitor, if you had a surface that looked like that (see Fig. \ref{entrevista Id}), that’s where you had problems, because there is no conduction current, you might say, that crosses that surface, but clearly this wire generates fields, this was a problem.\\\textit{Alejandro}: Yes, and we make the correction there.\\\textit{Martín}: Of course, that is where the correction happens. The one there that seems to be Ampère’s law to me would work without the correction, I don’t know about you, but I don’t think this is the same case, because actually, due to the curve, the net intensity is the wire, and it passes through the wire and that’s it. This is a nice surface, let’s put it like that.\\\textit{Diana}: And in general, the displacement current, when it appeared, was negligible, except when there was no common current and when it was zero, it was no longer negligible, and it was very important.\\\textit{Alejandro}: It is when I don’t have current intensity that I look at the displacement current term. \\\textit{Interviewer}: Diana, do you think that the displacement is null or is negligible in this case in particular?\\\textit{Diana}: No, I’m more on the side of it being null.\\\textit{Interviewer}: Why do you say that it is null there?\\\textit{Diana}: Because I have the whole conductor, so all the current would go there.
\end{quote}

\begin{figure}[htbp!]
\centerline{\includegraphics[width = 0.4\columnwidth]{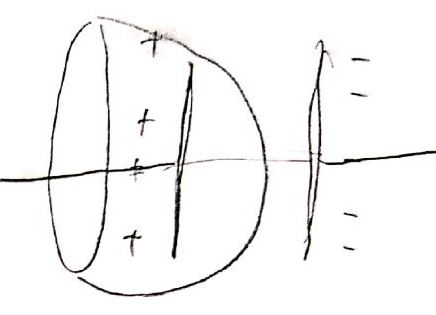}}
\caption{Hand-drawn sketch given by Martin during the interview.}
\label{entrevista Id}
\end{figure}

By attempting to predict the circulation of the magnetic field, the students firstly resort to comparing the Q2 situation with a ‘standard’ which is widely used in teaching and textbooks to introduce the concept of displacement current. Students frequently solve problems this way, but if the situations are not similar, it usually leads to conceptual and epistemic mistakes \cite{Ferguson1987,chi1989self}. In the answers from category B, as shown in the interview, the students do not manage to clearly connect the presence of the displacement current with a variable electric field. This reasoning leads them to conclude that the magnetic field circulation can be determined by using Ampère’s law with a conduction current which varies over time.

In category C (43,1\%), we group together the answers from students who consider the conduction current to be constant. We identified two sub-categories within this category: C1, where the students directly apply Ampère’s law to answer the question and C2, where they test ‘ad-hoc’ explanations which indicate a lack of comprehension of Ampère’s law and Ampère-Maxwell’s law. Let’s look at examples from each sub-category.

\begin{quote}
\textit{C1 - “I hope to get a result that is equal to what was predicted by Ampère’s law as the problem occurs when the surface does not include the current.”} (UPV/EHU-29)
\end{quote}

\begin{quote}
\textit{C2 - “Greater, because as there are more factors in the formula (surface), the product is greater.”} (UPV/EHU-1)
\end{quote}

The students in this category reveal a superficial knowledge of Ampère's and Ampère-Maxwell's laws. In particular, those in category C1 tend to apply Ampère's law by rote, evidencing a lack of understanding of both the validity framework of the law and the underlying physical phenomenon during the discharge of the spheres. On the other hand, students in category C2, as is the case of UPV/EHU-1, present explanations based on superficial approaches, such as mentioning more factors in the formula, showing a lack of understanding of the laws involved.

\subsection{\label{SectionIVC}Question 3}

Five categories emerge from analysing the answers. Table \ref{Tabla cuestión 3} shows the description of each one and their prevalence rates. 

\begin{table*}
 \caption{\label{Tabla cuestión 3} Answer categories and their percentages for question 3.}
\begin{tabular}{|c|p{12,9 cm}|c|}  
\hline
\textit{Category}
 & \textit{Description of the category} & \textit{\%}\\ \hline
A & Explanations which recognise the presence of a time-varying electric field associated with the magnetic field. & 9,2 \\ \hline
B & Explanations that associate the presence of the magnetic field with a static electric field. & 9,2 \\ \hline
C & Explanations based on poorly applied rote learning.& 66,2 \\ \hline
D & Incoherent. & 9,2 \\ \hline
E & They do not answer. & 6,2 \\ \hline
\end{tabular}
\end{table*}

The explanations included in category A (9.2\%) are based on arguments that recognise the presence of a variable electric field associated with the magnetic field demonstrated in the question. We have transcribed two typical answers below.

\begin{quote}
\textit{A - “There is a variable electric field pointing out of the page.”} (UDELAR-6)
\end{quote}

\begin{quote}
\textit{A - “This might be due to the presence of an electric field which varies over time, as this brings about a magnetic field.”}  (UPV/EHU-30)
\end{quote}

In category B (9,2\%), we include the answers with an incomplete level of comprehension, as they associate this magnetic field with an electric field without considering its variation. Below, we present examples to illustrate this category.

\begin{quote}
\textit{B - “I would explain it through the existence of an electric field, perpendicular to the magnetic field, pointing upwards. In this way, the magnetic field is displaced by the electric field to the left, in a circle.”} (UPV/EHU-6)
\end{quote}

\begin{quote}
\textit{B - “This magnetic field could have been created thanks to the existence of an electric field and a conduction current.”} (UPV/EHU-13)
\end{quote}

Category C (66,2\%) includes the students’ reasoning when they associate the magnetic field with conduction currents. The explanations are based on functional fixedness  \cite{viennot1996raisonner} on a strategy that is repeatedly applied in class for a specific case, associating the magnetic field with the electric current. It should be highlighted that two out of three answers fall into this category. Two typical answers are presented below to illustrate this category.

\begin{quote}
\textit{C - “In the centre of these lines, there would be a conductor cable, through which an upward intensity is circulating. This current generates the concentric magnetic field observed.”} (UPV/EHU-35)
\end{quote}

\begin{quote}
\textit{C - “The magnetic field occurs whenever there are charges in movement. As more charge is put in more movement, the magnitude of the magnetic field grows.”} (UPV/EHU-9)
\end{quote}

The students whose reasoning falls into this category ignore the fact that the magnetic field described in the question is in a empty region and they associate it incorrectly with conduction currents. This observation is demonstrated particularly clearly during the interviews.

\begin{quote}
    \textit{Interviewer}: How would you explain the presence of the magnetic field?\\\textit{Cintia}: With a conductor, right?\\\textit{Ricardo}: Yes, like that. A little wire.\\\textit{Cintia}: A straight conductor that passes through the centre.\\\textit{Ricardo}: Yes. \\\textit{Cintia}: Because always, even if you use the right-hand rule, due to the current that is circulating, you can see the direction of the field as well.\\\textit{Interviewer}: Can you think of anything else?\\\textit{Ricardo}: Well, you might have a field that varies over time, that for practical purposes acts like a current passing through it.\\\textit{Interviewer}: What field? Magnetic or electric?\\\textit{Ricardo}: Electric. You’d have two plates or two little spheres, and a field that varies over time. But what happens is that the mechanism would be a bit strange, but I think it would be the same. 
\end{quote}

The answers from Cintia and Ricardo show that they directly thought of a straight conductor which crosses the centre of the field lines, without considering that the magnetic field is in an empty region. The answers indicate a fixedness on associating the magnetic field with the conduction current. This shows how the students’ ways of thinking, both in interviews and in the written tests, consistently reflect reasoning based on functional fixedness associated with teaching that highlights specific cases (electric current in a very long straight wire) but do not relate the specific cases to the general theory \cite{viennot1996raisonner}. Only when other possibilities are raised in the interviews does the presence of a variable electric field associated with a charging capacitor emerge as a possible answer.

\subsection{\label{SectionIVD}Question 4}

Four explanatory categories emerge from analysing the students’ answers. Table \ref{Tabla cuestión 4} presents the different categories plus their respective percentages. 

\begin{table*}
 \caption{\label{Tabla cuestión 4} Answer categories and their percentages for question 4.}
\begin{tabular}{|c|p{12,5 cm}|c|}  
\hline
\textit{Category}
 & \textit{Description of the category} & \textit{\%}\\ \hline
A & Explanations that understand and correctly apply Ampère-Maxwell’s law to relate the concepts of magnetic field circulation and variation of electric flux. & 29,2 \\ \hline
B & Explanations that do not properly understand Ampère-Maxwell’s law and establish a relationship of proportionality between the electric field and the magnetic field. & 30,8 \\ \hline
B1 & They suppose that the proportionality constant is positive. & 21,6 \\ \hline
B2 & They suppose that the proportionality constant is negative.  & 9,2 \\ \hline
C & Incoherent. & 9,2 \\ \hline
D & They do not answer. & 30,8 \\ \hline
\end{tabular}
\end{table*}

In category A (29,2\%), we include the answers from students who recognise, according to Ampère-Maxwell’s law, that the magnetic field circulation is directly proportional to the rate of change of the electric field. A representative example of this category is presented below.

\begin{quote}
\textit{A – “When the electric field does not vary, there is no magnetic field. 
$$
\oint \vec{B} \cdot \overrightarrow{d l}=\mu_{0}\left(I_{c}+\varepsilon_{0} \frac{d \Phi_{E}}{d t}\right)
$$
When the electric field increases, the difference in electric flux over this area is a positive constant, as the increase is uniform. Similarly, when the electric field decreases, it is a negative constant. When the field is constant, the flux is also constant, as the surface area does not vary and therefore $\oint \vec{B} \cdot \overrightarrow{d l}=0$. $I_{c}=0$ in any case.”} (UPV/EHU-29)
\end{quote}

\begin{figure}[htbp!]
\centerline{\includegraphics[width = 0.5\columnwidth]{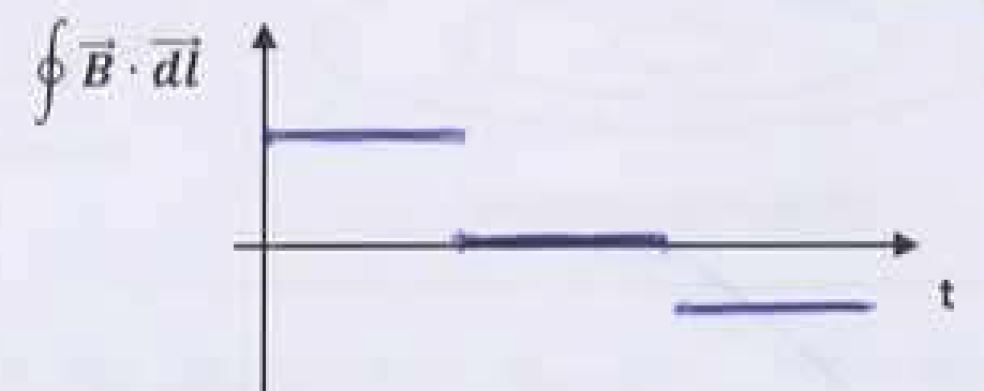}}
\caption{Hand-drawn sketch given by student UPV/EHU-29.}
\label{Gráfico UPV29}
\end{figure}

Within category A, we also include answers from a small group of students who plotted the magnetic field circulation over time, inversely to the correct answer. 

In category B (30,8\%), we group together the answers of students who do not manage to establish a connection between the magnetic field circulation and the variation of the electric field over time. These students suppose a directly proportional relationship between the electric field and the magnetic field. Approximately one in every three answers from this category explicitly indicates that the magnetic field opposes the electric field. Based on these results, we divide category B into two subcategories: B1, where the students suppose a relationship of proportionality between the fields, and B2, where they explicitly describe that these magnitudes have opposing signs. We present examples of each sub-category below.

\begin{quote}
\textit{B1 – “If the electric field module increases, by substituting $\oint \vec{B} \cdot \overrightarrow{d l}$ it also increases: the same as if it is decreasing and we substitute in the formula, its value drops.”} (UPV/EHU-1)
\end{quote}

\begin{figure}[htbp!]
\centerline{\includegraphics[width = 0.5\columnwidth]{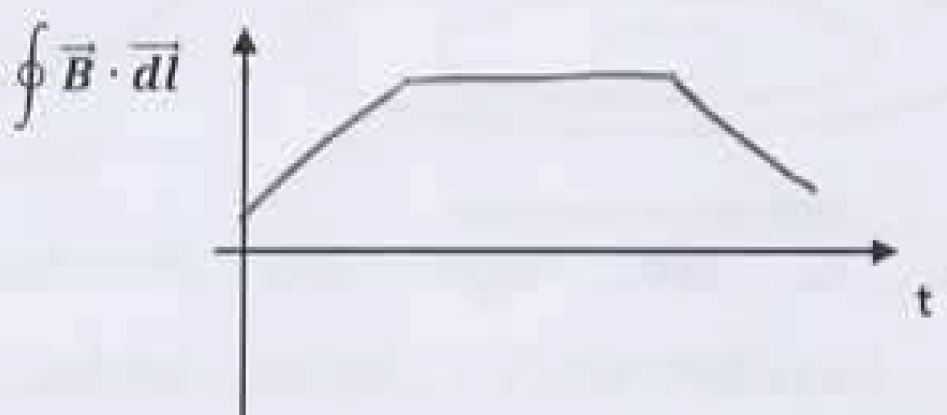}}
\caption{Hand-drawn sketch given by student UPV/EHU-01.}
\label{Gráfico UPV/EHU-1}
\end{figure}

\begin{quote}
\textit{B1 – “It acted like an integral. If there is a current in the direction of the arrow on the closed curve C, there would be a magnetic field in the same direction as C and, I have done a sketch considering that B and E are proportional.”} (UPV/EHU-18)
\end{quote}

\begin{figure}[htbp!]
\centerline{\includegraphics[width = 0.5\columnwidth]{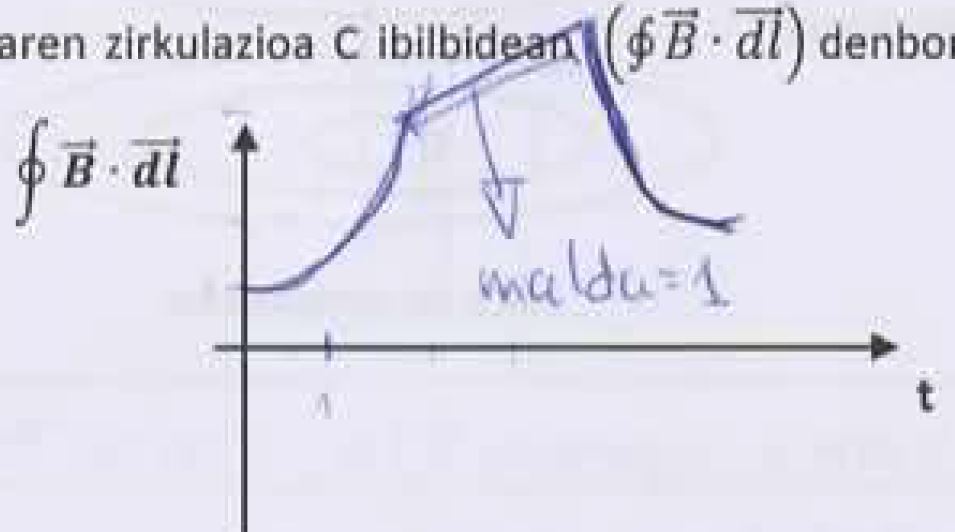}}
\caption{Hand-drawn sketch given by student UPV/EHU-18.}
\label{Gráfico UPV/EHU-18}
\end{figure}

\begin{quote}
\textit{B1 – “When the electric field is constant, there will be no change of flux and no magnetic field will be generated. When the electric field increases over time, the magnetic field will increase proportionally (E=Bc).”} (UPV/EHU–32)
\end{quote}

\begin{figure}[htbp!]
\centerline{\includegraphics[width = 0.5\columnwidth]{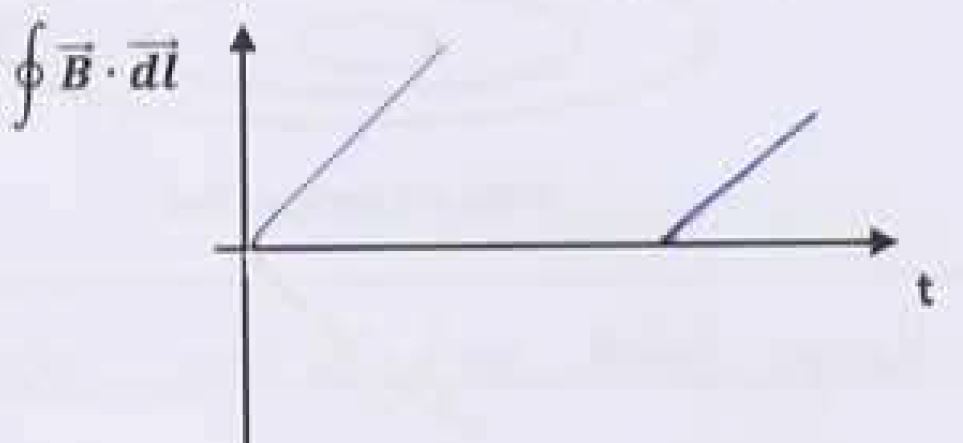}}
\caption{Hand-drawn sketch given by student UPV/EHU-32.}
\label{Gráfico UPV/EHU-32}
\end{figure}

\begin{quote}
\textit{B2 – “What the magnetic field does is try to compensate the change of the electric field lines. This means that if the module of $\vec{E}$ increases, $\vec{B}$ will drop and vice versa. And if $\vec{E}$ stays the same $\vec{B}$ will also remain the same.”} (UPV/EHU–10)
\end{quote}

\begin{figure}[htbp!]
\centerline{\includegraphics[width = 0.5\columnwidth]{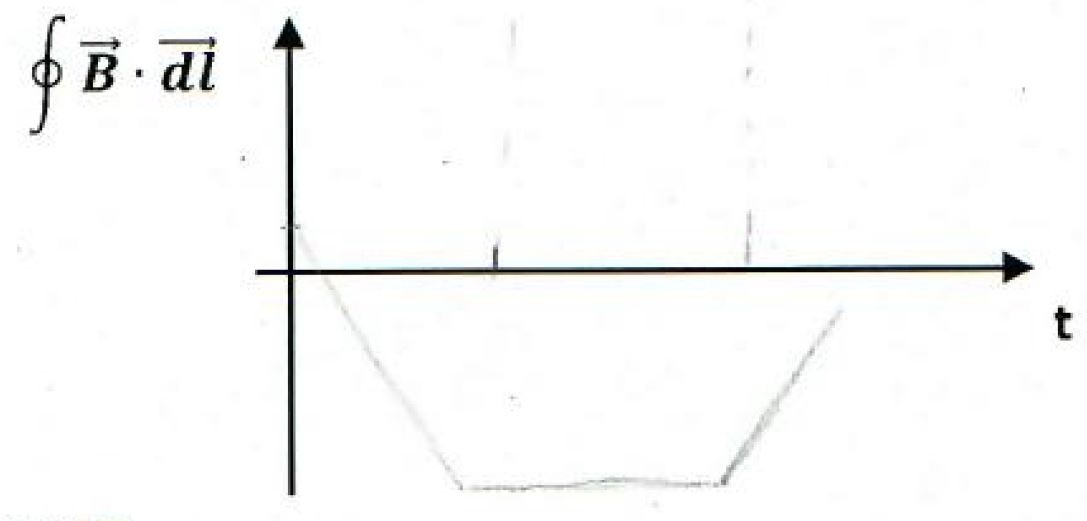}}
\caption{Hand-drawn sketch given by student UPV/EHU-10.}
\label{Gráfico UPV/EHU-10}
\end{figure}

From the answers provided by the students in category B, we can deduce that they believe the electric and magnetic fields to be directly proportional. This type of reasoning leads students to graphically represent the magnetic field circulation over time, with similar or opposing forms to the electric field graph. We also find answers such as from UPV/EHU-18 that include curved sections in their graphs. Although at first glance this might infer that this type of graph implies considering another type of relationship between the electric and magnetic fields, when analysing the explanations provided, we see that the difference is because, assuming the proportionality between the fields, they integrate the electric field to obtain the magnetic field circulation. In category B2, the UPV/EHU-10 student incorrectly applies an ‘ad-hoc’ version of Lenz’s rule, by claiming that the magnetic field is opposed to the change in the electric field. 

In general, the students do not provide explicit arguments in their answers to back their idea that the fields are proportional. In some specific cases, we identified answers which mention different equations connecting the electric and magnetic fields in different contexts, such as the example of the UPV/EHU–32 student who alludes to the relationship between these fields in the context of an electromagnetic wave. This type of rote-learning reasoning is also clear in the interviews. Here is an extract from one of them.

\begin{quote}
   \textit{Karina}: I imagine that by having the magnetic field circulation there, if the electric field varies, the magnetic field will have to vary in the same way. However, as these fields are gradients of each other, if one increases, the other would have to decrease. And this part (referring to the central section), as it remains the same over time, will be zero. 

\begin{figure}[htbp!]
\centerline{\includegraphics[width = 0.5\columnwidth]{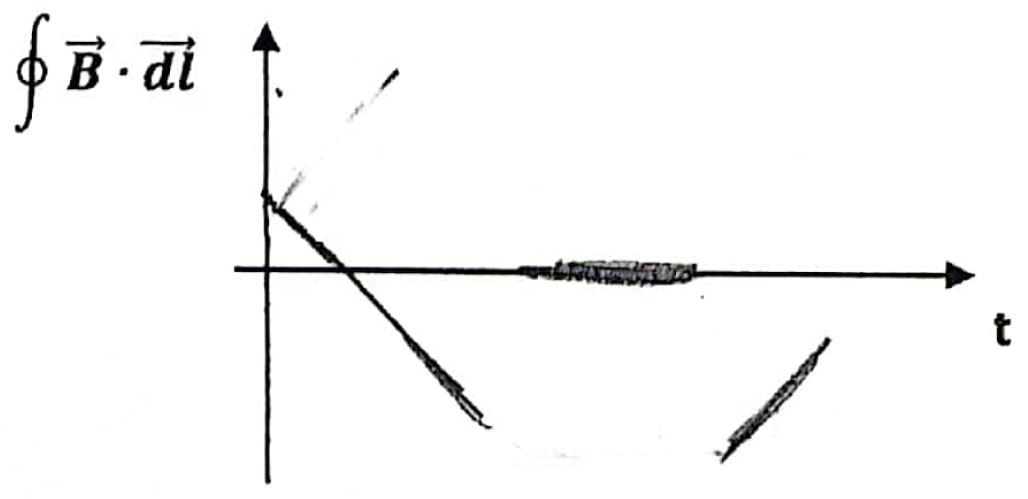}}
\caption{Hand-drawn sketch given by Karina during the interview.}
\label{Entrevista Karina}
\end{figure}
   
   \textit{Interviewer}: What do you mean when you say that a field is gradient of another?\\\textit{Karina}: That the fields are tied to equations, what were they called, partial derivatives, meaning that the fields are equipotential. If the electric field increases over time, the magnetic field will have to decrease over time. In an electric field variation, there would have to be a magnetic field.
 \end{quote}

We can see that Karina is working from the premise that the fields are proportional to build the graph for magnetic field circulation over time. When justifying her reasoning, she mentions a possible mathematical relationship between the fields based on what she has remembered and poorly assimilated from rote-learning, referring to a relationship between derivatives. Karina concludes that if a field undergoes changes, the other field should also change.

\section{\label{SectionV}Discussion of results} 

In this section, we discuss the results in relation to the research questions. Firstly, we analyse the students’ understanding when they face situations where Ampère’s law is not valid (RQ1). Next, we analyse how they understand and apply Ampère-Maxwell’s law (RQ2).  

Analysis of the students’ reasoning regarding the first research question reveals that the majority find it hard to interpret the concept of circulation and the term of current in Ampère’s Law (LO1), and to recognise its validity framework (LO2). In particular, when tackling question 1, barely one fifth of the students manage to identify the current which crosses different surfaces (LO1) and recognise that Ampère’s law produces contradictory results (LO2). Introducing the curved surface generates different answers regarding the magnetic field circulation, demonstrating a lack of comprehension of this operator and Ampère’s law among around two thirds of the students (categories B and C). We identified several explanations that show a limited or incorrect comprehension of Ampère’s law, highlighting the confusion between flux and circulation for approximately one in every four students (categories B2 and C2). This difficulty converges with results from the study by Hernández et al. \cite{Hernandez2023}. We also noted incorrect reasoning based on the current among students from categories B1 and C1, where some mistakenly conclude that if the current which crosses a surface is null, then the magnetic field is also null, reflecting simple causal reasoning \cite{halbwachs1971reflexions,Leniz2017}. This type of reasoning has been documented in the literature in a different context \cite{wallace2010,Campos2023}. Furthermore, some students incorrectly suppose that the current which must be considered to calculate the magnetic field circulation is the current enclosed by the curve and not the current which crosses a surface bound by it. This might be because the net current is defined in this way by some textbooks \cite{young2019university,walker2014halliday}.

When analysing the students’ reasoning in relation to the application of Ampère-Maxwell’s law (RQ2), we find that the majority find it hard to recognise the existence of the displacement current and its role in Ampère-Maxwell’s law (LO3). They also find it hard to relate the magnetic fields to time-varying electric fields (LO4) and understand the relation between the rate of change of the electric field  and the magnetic field circulation (LO5). In particular, for learning objective LO3, we find that in question 2, around 10\% of the students manage to identify the displacement current and correctly apply Ampère-Maxwell’s law. In this question, we observe that more than half of the students (categories B and C) do not manage to identify the displacement current. Our results suggest that this might be attributed to incomplete comprehension of the relationship between the displacement current and the time-varying electric field, and functional fixedness \cite{viennot1996raisonner} on the appearance of the displacement current associated with the standard example of a charging capacitor. This fixedness on a strategy prevents students from recognising the presence of this current in other contexts, leading them to suppose that Ampère’s law is still valid.

Regarding LO4, we can see that approximately 10\% of students associate the magnetic field with a time-varying electric field in question 3. We also find that a similar number of students mistakenly attribute the generation of magnetic fields to stationary electric fields. This result converges with findings from other research which shows that the students must consider the electric fields as sources of magnetic fields \cite{guisasola2004difficulties}. However, almost two out of three students maintain that the existence of a magnetic field in the empty region is due to the conduction currents in the field lines’ axis of symmetry. This type of reasoning reveals the presence of a functional fixedness  \cite{viennot1996raisonner} on specific case studies, where it is repeatedly taught that if there is a magnetic field in a particular place, this is because there is an electric current.

Regarding LO5, in question 4, we find that around one third of the students understand the meaning of the concepts of magnetic field circulation and variation of the electric flux, and the relationship between the two in Ampère-Maxwell’s law. However, we note that almost one third of the students (category B) argue that the electric and magnetic fields are directly proportional. One possible explanation of this type of reasoning could be attributed to the confusion between the electric field and its time derivative. The confusion between a magnitude and its derivative is a widely-documented topic in the literature, due to the inherent complexity of derivatives \cite{Trowbridge1980,Trowbridge1981,allen2001investigation}. The literature also provides research on students’ ideas that the  emf is directly proportional to the magnetic field within the framework of Faraday’s law \cite{guisasola2013university}, although we have not found other studies suggesting that the students consider electric and magnetic fields are proportional to each other. In the same question, around 10\% of students (category B2) present arguments which incorrectly apply an ‘ad-hoc’ version of Lenz’s rule. These answers reveal incomplete understanding on how magnetic field circulation is related to temporary variation of the electric field. The literature explains that this reasoning might be because, when two topics are perceived as similar, the learning from one might interfere with the other \cite{Heckler2010}. 
Finally, we note that both in this question and in question 2  more than 25\% of the students fail to provide an answer. These results are significantly higher than the percentage of students failing to answer questions 1 and 3, suggesting a lack of knowledge about the relationship between the time-varying electric field and magnetic field circulation. This could be originated in the lack of attention and practice in the application of the Ampère-Maxwell law, an aspect reflected in the concise treatment 
of this law in textbooks and the small number of proposed problems compared to other fundamental laws \cite{Bagno1997,Kesonen_2011}.

\section{\label{SectionVI}Conclusion} 

In this study, we researched the learning difficulties faced by students when applying Ampère-Maxwell’s law. To carry out this research, we used phenomenography to analyse the written responses from 65 undergraduate students to four questions on Ampère and Ampère-Maxwell’s laws. This methodology helped us to identify the different types of reasoning. To further understand how students think, we completed our research by interviewing twelve students, exploring the same questions as posed in the study. 

Regarding the students’ learning difficulties when tackling situations where Ampère’s law is not valid (RQ1), we noted that the majority present inappropriate comprehension of the concept of circulation. They face difficulties to interpret the relationship between the terms in Ampère’s law to recognise the current crossing a curved surface, which leads to a lack of comprehension regarding the validity framework of this law. These findings highlight the need to explicitly define the concept of circulation before binding it to Ampère’s law. Although this operator is difficult to understand, experimental activities can be designed \cite{Leclerc} which allow students to conceptualise it more solidly. In addition, it is fundamental to address classroom situations where the students must apply Ampère’s law to different surfaces bound by the same curve, to develop appropriate comprehension of each term and its validity framework.

Regarding the students’ learning difficulties when facing situations where they must apply Ampère-Maxwell’s law (RQ2), we identified that they find it challenging to recognise the relationship of both the magnetic field and magnetic field circulation with a time-varying electric field. This limited comprehension means that students find it hard to recognise situations where the displacement current appears, and to mistakenly argue that a magnetic field can only be associated with an electric conduction current. These results highlight the need to address a variety of situations in the classroom, beyond the specific case of an RC circuit, which make it easier to recognise the presence of the displacement current and its relationship with the variable electric fields \cite{frish1957,feynman1963feynman,suarez2022,suarezEJPE}.

The main limitation of this work revolves around the number of students who take part in the study and the methodological focus based on qualitative analysis. Despite these limitations, the consistency of the difficulties identified in the written tests and the interviews suggests that our findings are robust, which could be confirmed in larger-scale studies using quantitative methods. Furthermore, several of the conceptual difficulties coincide with previous findings in different contexts.

We believe that it is fundamental to carry out more research on students’ learning difficulties when applying Ampère-Maxwell’s law. This would not only broaden our knowledge in the area but would also help to develop a more complete corpus to design future teaching materials. In our next study, we shall design a teaching-learning sequence that could prevent the appearance of some previously-reported learning difficulties. Working from our findings, we shall consider a more appropriate working hypothesis to be that, in future sequences, phenomena should be addressed where the displacement current appears and Ampère-Maxwell’s law is analysed, before studying electromagnetic induction and Faraday’s law. We think that introducing the displacement current after analysing and discussing the validity framework for Ampère’s law, might help to address difficulties faced by students more effectively.  Furthermore, by studying Ampère-Maxwell’s law first, this would simplify tackling Faraday’s law and Lenz’s rule subsequently, as the relationship between the magnetic field circulation and the rate of change of the electric field is not negative.

\section*{Acknowledgment}
The research that gives rise to the results presented in this publication received funds from the Agencia Nacional de Investigación (ANII), Uruguay, under the code POS\_NAC\_2023\_5\_177677. 
The authors would like to thank PEDECIBA (MEC, UdelaR, Uruguay).  Part of this research was funded by the Spanish government (MINECO\textbackslash FEDER PID2019 -105172RB-I00)

%\nocite{*}

\appendix

\section{\label{apppendix}Questionnaire} 

\textbf{Question 1.} Figure \ref{A1} shows two electrically-charged conducting spheres with opposite charges. By joining the spheres using a conducting wire, a conduction current circulates until the spheres have the same potential. Consider the closed curve \textit{C} and the surfaces $S_{1}$ and $S_{2}$, bound by this curve.
How would you apply Ampère’s law to surface S1 and to surface S2? 

\begin{figure}[h!]
\centerline{\includegraphics[width = 0.5\columnwidth]{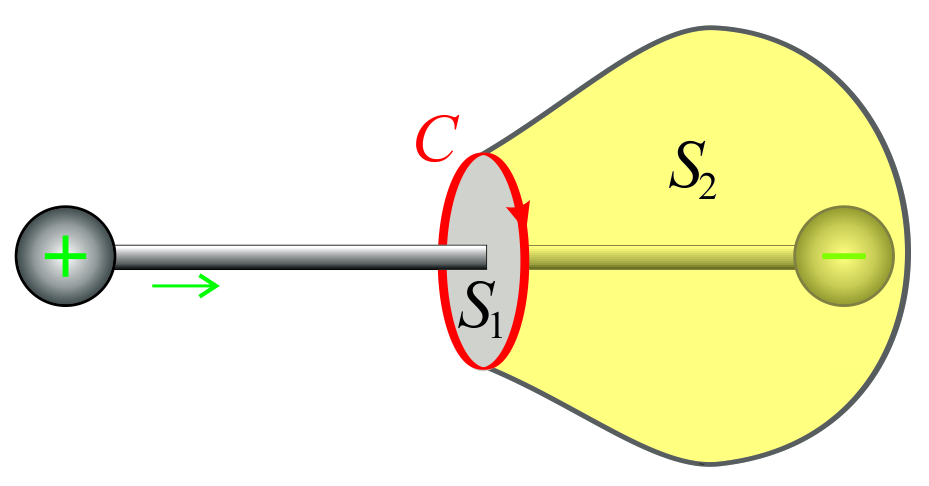}}
\caption{Two charged conducting spheres with opposite charges, connected using a conducting wire, a closed curve \textit{C} and two surfaces $S_{1}$ and $S_{2}$ that the curve borders.}
\label{A1}
\end{figure}

\textbf{Question 2.} Figure \ref{A2} shows two electrically-charged conducting spheres with opposite charges. By joining the spheres using a conducting wire, a conduction current circulates through the wire until the spheres have the same potential. 
Consider the closed curve \textit{C} and the surface \textit{S} that it borders. If you apply Ampère-Maxwell’s law to calculate the magnetic field circulation ($\oint \vec{B} \cdot \overrightarrow{d l}$) along curve \textit{C}, would you expect to obtain a result that is greater, equal to or lower than what was predicted by Ampère’s law? 

\begin{figure}[h!]
\centerline{\includegraphics[width = 0.4\columnwidth]{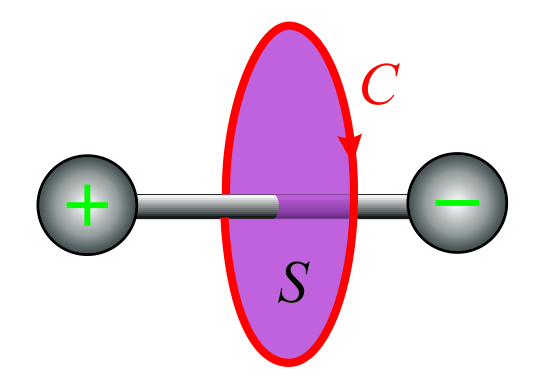}}
\caption{Two conducting spheres charged with opposite charges and connected to each other by a conducting wire, a closed curve \textit{C} and a bounding surface.}
\label{A2}
\end{figure}

\textbf{Question 3.} A magnetic field is detected in a empty region and its field lines form concentric circumferences as indicated in Fig. \ref{A3}. 
How would you explain the presence of this magnetic field? 

\begin{figure}[h!]
\centerline{\includegraphics[width = 0.6\columnwidth]{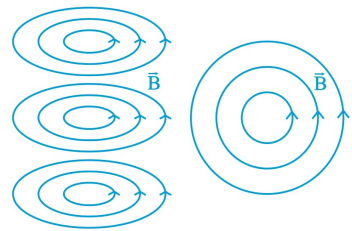}}
\caption{A perspective view of the magnetic field lines is shown on the left and a top view on the right.}
\label{A3}
\end{figure}

\textbf{Question 4.} Consider an electric field pointing into the page and a closed curve \textit{C}. The electric field module varies over time as shown in the graph in Fig. \ref{A4}. 
Sketch the graph of the magnetic field circulation ($\oint \vec{B} \cdot \overrightarrow{d l}$) along curve C over time.

\begin{figure}[h!]
\centerline{\includegraphics[width = 0.8\columnwidth]{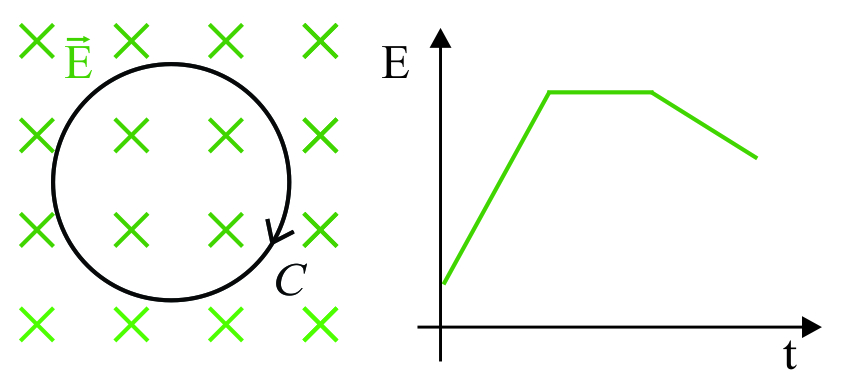}}
\caption{On the left is shown an electric field pointing into the page and a closed curve C. On the right is the graph of the electric field as a function of time.}
\label{A4}
\end{figure}

\bibliography{referencias} %Produces the bibliography via BibTeX.

\providecommand{\noopsort}[1]{}\providecommand{\singleletter}[1]{#1}%

\end{document}